\documentstyle[12pt,epsf]{article}
\begin{document}
\title{The 1-soliton in the $SO(3)$ gauged Skyrme model with mass term}
\author{
{\large Y. Brihaye}$^{\diamond}$,
{\large J. Burzlaff}$^{\ddagger\star}$,
{\large V. Paturyan}$^{\dagger}$,
and {\large D. H. Tchrakian}$^{\dagger \star}$ \\ \\
$^{\diamond}${\small Physique-Math\'ematique, Universit\'e de
Mons-Hainaut, Mons, Belgium}\\ \\
$^{\ddagger}${\small School of Mathematical Sciences, Dublin
City University, Dublin 9, Ireland}\\ \\
$^{\dagger}${\small Department of
Mathematical Physics, National University of Ireland Maynooth,} \\
{\small Maynooth, Ireland} \\ \\
$^{\star}${\small School of Theoretical Physics -- DIAS, 10 Burlington Road,
Dublin 4, Ireland }}

\date{}
\newcommand{\dd}{\mbox{d}}
\newcommand{\tr}{\mbox{tr}}
\newcommand{\la}{\lambda}
\newcommand{\ka}{\kappa}
\newcommand{\al}{\alpha}
\newcommand{\ga}{\gamma}
\newcommand{\de}{\delta}
\newcommand{\si}{\sigma}
\newcommand{\bomega}{\mbox{\boldmath $\omega$}}
\newcommand{\bsi}{\mbox{\boldmath $\sigma$}}
\newcommand{\bchi}{\mbox{\boldmath $\chi$}}
\newcommand{\bal}{\mbox{\boldmath $\alpha$}}
\newcommand{\bpsi}{\mbox{\boldmath $\psi$}}
\newcommand{\brho}{\mbox{\boldmath $\varrho$}}
\newcommand{\beps}{\mbox{\boldmath $\varepsilon$}}
\newcommand{\bxi}{\mbox{\boldmath $\xi$}}
\newcommand{\bbeta}{\mbox{\boldmath $\beta$}}
\newcommand{\ee}{\end{equation}}
\newcommand{\eea}{\end{eqnarray}}
\newcommand{\be}{\begin{equation}}
\newcommand{\bea}{\begin{eqnarray}}
\newcommand{\ii}{\mbox{i}}\newcommand{\e}{\mbox{e}}
\newcommand{\pa}{\partial}\newcommand{\Om}{\Omega}
\newcommand{\vep}{\varepsilon}
\newcommand{\bfph}{{\bf \phi}}
\newcommand{\lm}{\lambda}
\def\theequation{\arabic{equation}}
\renewcommand{\thefootnote}{\fnsymbol{footnote}}
\newcommand{\re}[1]{(\ref{#1})}
\newcommand{\R}{{\rm I \hspace{-0.52ex} R}}
\newcommand{\N}{{\sf N\hspace*{-1.0ex}\rule{0.15ex}%
{1.3ex}\hspace*{1.0ex}}}
\newcommand{\Q}{{\sf Q\hspace*{-1.1ex}\rule{0.15ex}%
{1.5ex}\hspace*{1.1ex}}}
\newcommand{\C}{{\sf C\hspace*{-0.9ex}\rule{0.15ex}%
{1.3ex}\hspace*{0.9ex}}}
\newcommand{\eins}{1\hspace{-0.56ex}{\rm I}}
\renewcommand{\thefootnote}{\arabic{footnote}}

\maketitle
\begin{abstract}
The solitons of the $SO(3)$ gauged Skyrme model with no pion-mass
potential were studied in Refs.~\cite{nl,jmp}. Here, the effects of the
inclusion of this potential are studied. In contrast with the (ungauged)
Skyrme model, where the effect of this potential on the solitons is
marginal, here it turns out to be decisive, resulting in very different
dependence of the energy as a function of the Skyrme coupling constant.
\end{abstract}
\medskip
\medskip
\newpage

\section{Introduction}
\label{introduction}

The solitons of the $SO(3)$ gauged Skyrme model with no pion mass
potental were studied in detail numerically in Refs.~\cite{nl,jmp}.
The model
\be
\label{L}
{\cal H}_0=-\frac{1}{4}{\mbox Tr}|F_{ij}|^2+
{\mbox Tr}\left((U^{-1}D_iU)^2+\kappa ([U^{-1}D_iU,U^{-1}D_jU])^2\right)
\ee
was determined by
the vector gauging prescription defined by the covariant derivative
\be
\label{gauge}
D_iU=\pa_iU+[A_i,U]\ ,\qquad U=\cos f+i{\bf n}.\bsi\sin f\ ,
\ee
where $U$ is an element of $SU(2)$, and the (antihermitian) Yang-Mills
connection and curvature are in the algebra.

Replacing $\pa_iU$ in the usual Skyrme model~\cite{S} by \re{gauge}
yields the model \re{L} studied in \cite{nl,jmp}. In the present work, we
augment this system ${\cal H}_0$ by the familiar pion-mass\footnote{Other
potentials can also be employed if exponential localisation is not
insisted on. In particular for solitons with nonvanishing magnetic
flux~\cite{BHT} exponential localisation is not obtained with the
pion-mass potential} potential
\be
\label{pot}
V=\la (1-\cos f)
\ee
such that the system we study here is
${\cal H}={\cal H}_0+\la (1-\cos f)$.

The effects of this,
as we shall see from our numerical work in Section~\ref{numerics}, are
considerable, in contrast with the usual (ungauged) Skyrme model~\cite{S}
where the only effect is the quantitative one of rendering the asymptotic
decay exponential rather than power like.

We are interested in
{\it unit} Baryon number solitons, hence only in spherically symmetric
solutions. Imposition of spherical symmetry leads to
\be
A_i=-\frac{i}{2}\left(\frac{a(r)-1}{r}\right)\vep_{ijk} \si_j\hat
x_k\ , \quad n_i=\hat x_i\, \quad f=f(r)\ . \ee The static
Hamiltonian reduces to the one dimensional energy density
functional
\bea H = 4(a')^2 &+& {2(a^2- 1)^2\over{r^2}} +
\frac{1}{2}[r^2(f')^2 + 2a^2 \sin^2f] \nonumber \\ &+& 4\kappa a^2
\sin^2f\left[(f')^2 + a^2{\sin^2f\over {2 r^2}}\right]+\la (1-\cos
f)r^2\ , \label{H}
\eea
whose equations of motion are to be solved
subject to the asymptotic conditions
\bea
\lim_{r\to 0}f(r)&=&\pi\ ,\qquad \lim_{r\to\infty}f(r)=0\ ,
\label{asymf}\\
\lim_{r\to 0}a(r)&=&1\ ,  \qquad \lim_{r\to\infty}a(r)=\pm1\ .
\label{asyma}
\eea
Recall that in the $\la=0$ case in \cite{nl,jmp}, the boundary values
\re{asyma} are instead
\be
\label{asyma0}
\lim_{r\to 0}a(r)=1\ ,  \qquad \lim_{r\to\infty}a(r)=0,\pm1\ .
\ee

The asymptotic values \re{asymf} of the function $f(r)$
are consistent with analyticity at the origin and finite energy at
infinity. The latter condition is ensured by the presence of the pion
mass potential \re{pot}. This contrasts with the ungauged model
(resulting from putting $a(r)=1$ in \re{H}) for which the asymptotic
condition \re{fasym} follows from the finite energy condition whether
or not $\la=0$. If in the gauged model one puts $\la=0$, then condition
\re{asymf} can be imposed as a constraint for {\it unit} Baryon charge,
so that the solutions studied in \cite{nl,jmp} are constrained solutions..

Concerning the asymptotics of the function $a(r)$,
the first member of \re{asyma} ensures differentiability at the origin,
but the second member does not follow from the requirement of finite
energy by a naive inspection of the functional \re{H}. Doing the latter
would yield instead the weaker condition
\be
\label{a'} \lim_{r\to\infty}a'=0\quad\Rightarrow\quad
\lim_{r\to\infty}a=a_0\in{\bf R}\ee fixing only the $r$ derivative
of $a(r)$ at infinity. The second members of \re{asyma} and \re{asyma0},
result from a careful asymptotic analysis of the second order equations
of motion to be given in Section~\ref{asymptotics}. The results of the
numerical analysis will be reported in Section~\ref{numerics}, and
summarised in Section~\ref{summary}.

\section{Asymptotic analysis: dominant balance}
\label{asymptotics}

The Euler-Lagrange equations for \re{H} are
\bea
(r^2 + 8 \kappa a^2 \sin^2 f )f'' + 2r f' + 16 \kappa a a' \sin^2
f f'+ 8 \kappa a^2 \sin f \cos f (f')^2
 && \nonumber \\ - 2 a^2 \sin f \cos f
- 8\kappa \frac{a^4 \sin^3 f \cos f}{r^2} - \la r^2 \sin f &=& 0\ ,
\label{eqf} \\
a'' + \frac{a(1-a^2)}{r^2} - \frac{1}{4}a\sin^2 f -
\kappa a \sin^2 f (f')^2 - \kappa \frac{a^3 \sin^4 f}{r^2} &=& 0\ ,
\label{eqa}
\eea
and the following boundary conditions are imposed:
\be
\label{22} f\rightarrow 0, \;\;\; a\rightarrow a_0 \in {\bf R},
\;\;\; f' \rightarrow 0, \;\;\; a' \rightarrow 0\;\;\; {\rm as}
\;\; r\rightarrow \infty\ .
\ee
We want to show that there are no
solutions to the differential equations \re{eqf}-\re{eqa} in the
asymptotic region if $a_0 = \pm 1$ is not satisfied. The $\la\neq 0$ case
of interest here differs from the case where $\la = 0$. First, here
we will find exponential rather than power decay of the function
$f$. Second, the asymptotic behaviour not ruled out by the
asymptotic analysis presented here is $a_0 = \pm 1$, whereas in
the case $\la =0$ it is $a_0 =0,\pm 1$. The numerical results are,
of course, in accordance with the asymptotic analysis.

To identify the dominant terms in the asymptotic region we use
that
\[
8 \kappa a^2 \sin^2 f f'' \ll r^2 f'', \;\;\;
16\kappa a a' \sin^2 f f' \ll 2r f',
\]
\[
8\kappa a^2 \sin f\cos f (f')^2 \ll 2r f', \;\;\; 8\kappa
\frac{a^4 \sin^3 f \cos f}{r^2} \ll 2a^2 \sin f\cos f\ll\la r^2 \sin f\ .
\]
Therefore to leading order \re{eqf} reduces to
\be
\label{24} r^2 f'' + 2r f' - \la r^2 f = 0\ ,
\ee
which can be solved
in terms of Bessel functions. The asymptotic behaviour of the
solution is
\be
\label{25} f\approx \frac{f_0}{r} \exp (-\sqrt{\la}r)\ .
\ee
Note that in the case $\la = 0$, equating the terms of leading order
in \re{eqf} yields
\be
\label{24a}
r^2 f'' + 2r f' - 2 a_0^2 f = 0\ .
\ee
This equation has
\be
f=f_0 r^{-(1+\sqrt{1+8a_0^2})/2}
\ee
as a family of solutions which satisfies the boundary condition for $f$.

In the asymptotic region we also have
\[
\kappa a \sin^2 f(f')^2 \ll \frac{1}{4}a \sin^2 f,\;\;\; \kappa
\frac{a^3 \sin^4 f}{r^2} \ll \frac{1}{4}a \sin^2 f\ .
\]
In terms of $A=a-a_0$ we therefore obtain from \re{eqa} to leading order
\be
\label{27} r^2 A'' + (1-3a_0^2)A = a_0 (a_0^2 -1) + \frac{r^2}{4}
(a_0 + A) f^2\ .
\ee
Because of the exponential decay of $f$ in the
case $\la\neq 0$, the equation reduces to
\be
\label{27a}
r^2 A'' + (1-3a_0^2)A = a_0 (a_0^2 -1)\ .
\ee

For $3a_0^2 = 1$ the general solution of \re{27a} is
\be
A= \frac{2}{3\sqrt{3}} \log r + c_1 + c_2r
\ee
which does not go to zero and is unacceptable.
The solutions of \re{27a} for $|a_0|<\frac{1}{2}$, $a_0=\pm\frac{1}{2}$
and $|a_0|>\frac{1}{2}$ are, respectively,
\bea
A&=& \frac{a_0^3 - a_0}{1-3a_0^2} +
\sqrt{r} \left[c_3 \cos(\frac{1}{2}\sqrt{3 - 12 a_0^2}\log r) +
c_4 \sin (\frac{1}{2}\sqrt{3-12 a_0^2}\log r)\right]\ ,
\label{sol1} \\
A&=& \mp\frac{3}{2} + c_5\sqrt{r} + c_6
\sqrt{r} \log r\ , \label{sol2} \\
A&=& \frac{a_0^3 - a_0}{1-3a_0^2} +
\sqrt{r} \left( c_7 r^{\sqrt{12 a_0^2 - 3}/2} + c_8 r^{-\sqrt{12
a_0^2 - 3}/2}\right)\ , \label{sol3}
\eea
which for $a_0\neq \pm 1$ do not go to zero either. So there is no
solution of the equation \re{eqa} for which $a_0\neq \pm 1$. Equation
\re{27a} has solutions with acceptable asymptotic behaviour only if
$a_0 =\pm 1$, given by \re{sol3}. In this case $A\approx c_8 r^{-1}$
(with $c_7=0$)is a solution with acceptable behaviour in the asymptotic
region.

The situation here should be contrasted with the case where $\la = 0$.
There \cite{nl}, for $a_0 =0$,
\be
f\approx \frac{f_0}{r}, \;\;\;\; A\approx A_0 r^{(1-\sqrt{f_0^2 -
3})/2}
\ee
is an acceptable solution of \re{24a} and \re{27} in the
asymptotic region if $|f_0|>2$. In all cases the existence of
acceptable solutions for small and large $r$ of course does not
guarantee the existence of a solution which has both, acceptable
behaviour for small and for large $r$.

\section{Numerical results}
\label{numerics}

Before presenting our numerical results for the $\la\neq0$ case, let us
recall those for the system \re{H} with $\la=0$, carried out in
\cite{nl,jmp}. With respect to the Skyrme coupling constant $\kappa$,
only two energy branches of stable solutions were found. These are
presented in Figure 1 and labelled $A$ and $B$ respectively. The branch $A$
corresponding to $a(\infty)=1$ starts from $\ka=0$ and persits up to a
cusp critical point $\ka_1^{cr}$. The classical energy is an increasing
function of $\ka$ (with $E(\ka=0)=0$). Branch $B$ corresponding to
$a(\infty)=0$ starts from another critical point $\ka_2^{cr}$ and
persists up to $\ka\to\infty$. Branches $A$ and $B$ are bridged by a
branch $A'$ which like $A$ has $a(\infty)=1$ and is known to be
unstable~\cite{KTZ}; the numerical values  $\kappa_1^{cr}\approx 0.8091$ , 
$\kappa_2^{cr}\approx 0.6914$ were found in \cite{jmp}.
No solutions with $a(\infty)=-1$
were found, for which we do not have an explanation.

As we have no explanation for our inability to construct solutions with
$a(\infty)=-1$ for the $\la=0$ system, we have 
used a different numerical procedure as well as the 
alternative ways to implement the boundary conditions to attempt to
construct them and/or to
recover the results of \cite{nl,jmp}. Namely, this involves the use
of the boundary values \re{asymf} for the chiral function $f(r)$, while
for the gauge function $a(r)$ we have used the first member of \re{asyma}
together with \re{a'}. In this way, we indeed recovered the branches
$A$, $A'$ and $B$ found in \cite{nl,jmp}  but nothing else.
This provides a good check on the
correctness of the numerical procedure to be applied to the
$\la\neq0$ system in the present work.


After all these tests involving the boundary condition \re{a'}
we have applied our numerical routines to the $\la\neq 0$
system and we now discuss the results.
Various types of solutions are presented below for the value of
$\la = 1.0$ but we have checked that for generic
values of this parameter, the corresponding solutions display the same
qualitative features . 

First, the numerical analysis confirms the occurence of a branch
of solutions obeying $a(0)=a(\infty)=1$. It exists for all values of
$\kappa$ and the energy increases monotonically with this parameter,
as illustrated on Fig. 2 (branch $A$). This solution can further be
characterized by $x_m$, the value of $x$ where $a(x)$ attains its local
minimum. The following results hold~:
\be
       {\rm lim}_{\kappa \rightarrow 0} x_m = 0 \ \ \ , \ \ \ 
       {\rm lim}_{\kappa \rightarrow \infty} x_m \equiv \tilde x_m
       \approx 1.33.
\ee
For sufficiently high values of $\kappa$, our numerical analysis also
reveals the existence of two branches of solutions obeying
$a(0)= -a(\infty)=1$. For instance, for $\kappa  > \kappa_{c}$
($\kappa_{c}\approx 2.297$ in the case $\la = 1$), 
we constructed two different branches of solutions, labelled $C$ an
$C'$ on Fig. 2. As indicated the classical energies of these solutions
are slightly higher than the corresponding one on branch $A$. In the limit
$\kappa \rightarrow \kappa_c$, the two branches $C$ and $C'$ coincide.

We also find (using obvious notations $E_A=$ energy on branch $A$,...)
\be
{\rm lim}_{\kappa \rightarrow \infty} (E_C - E_A) = 0  \ .
\ee
The two new solutions can further be characterized by the value $x_0$
where the function $a(x)$ takes its (unique, as far as we can see) node.
The evolution of $x_0$ as a function of $\kappa$ for the two branches is
displayed on Fig. 3; this figure clearly indicates that, for 
$\kappa$ fixed, the two solutions are distinguished  by the position of their
node. The numerical results suggest that
\be
{\rm lim}_{\kappa \rightarrow \infty} x_0 = 
{\rm lim}_{\kappa \rightarrow \infty} x_m 
= \tilde x_m  \ \ \ \ {\rm for \ branch \ C   } \ \ ,    
\ee
\be
{\rm lim}_{\kappa \rightarrow \infty} x_m 
= \infty \ \ \ \ {\rm for \ branch  \ C'   } \ \ ,    
\ee

Finally, on Fig. 4, the profiles of $a(x)$ for different values of
$\kappa$ and on the different branches are compared. It is seen that, 
on the interval $[0, x_m]$, the function $a(x)$
corresponding to the solutions $C$ and $C'$ deviates only a little from
the corresponding one  on branch $A$ . 
The differences between the three solutions are rather perceptible
on the interval $[x_m, \infty]$.

Quantitatively the value of $\la$ affects
the value of $\ka_c$ slightly. We find numerically $\kappa_c= 2.423\ ,\
2.359\ ,\ 2.297$ respectively, for  $\lambda = 0.1\ ,\ 0.5\ ,\ 1.00$.
For all cases, we checked that the profiles of $a(x)$ and of $f(x)$ obey
the asymptotic behaviours obtained in the previous section, providing
a nice check of our numerics.
The various solutions were obtained by using the 
subroutine COLSYS \cite{colsys} (based on the damped Newton method
of quasi-linearization, a brief description of it is presented in
the Appendix of \cite{bhk}). 
The solutions were obtained with a high degree of accuracy~: typically with
errors less than $10^{-8}$.

The solutions displayed by Fig. 2, constitute the main result of the
present work. Comparison of Figs. 2 and 1 shows that the inclusion of
the pion-mass potential in the gauged sigma model has a considerable
qualitative effect compared to the case when this potential is absent.

Concerning the stability of the pattern of solutions displayed in Fig. 2,
we have not carried out a detailed quantitative analysis. However we
beleive that the solutions on the branches $A$, $C$  and $C'$ possess
respectively zero, one and two unstable modes. The stability of branch
$A$ is guaranteed by the toplogical lower bound associated with the
nontrivial asymptotics of the Skyrme field. 
The plot of the energy corresponding to the branches $C$, $C'$
terminates into a cusp, typical of catastrophe theory (see e.g. 
\cite{kus}). In such situations it is believed (and it was
demonstrated numerically in a particular case \cite{bks}) 
that the number of negative modes of 
the upper branch of the cusp exceed by one unit the number of negative
modes of the lower branch.
With our expectation of the number of unstable modes,
the calculation of the Morse index 
\be
\label{morse}
   \xi \equiv \sum_q (-1)^q N_q
\ee
(the sum runs over the classical solutions, $q$ counts the number
of negative modes of the solution and $N_q$ represents the number of them) leads to  $\xi = 1$
irrespectively of the parameter $\ka$.

\section{Summary}
\label{summary}

We have investigated the effect of adding the pion mass-potential,
namely the last term in \re{H}, to the Skyrme model. This potential
enforces the conventional asymptotic values \re{asymf} of the chiral
function $f(r)$, but in the gauge-decoupled case, with $a(r)=1$ \re{H},
\re{asymf} is independently guaranteed by the finite energy condition.
Not surprisingly in that case~\cite{S}, with $a(r)=1$, switching $\la$
on or off results in no appreciable qualitative change in the Skyrme
soliton. The power dacay for the $\la=0$ case is replaced by an
exponential decay in the $\la\neq 0$ case.

The situation is quite different in the gauged case, where the potential
is necessary to have \re{asymf}. On the other hand
imposition of \re{asymf} also guarantees {\it unit} Baryon charge, hence
it is tantamount to insisting on {\it unit} Baryon
charge even if $\la=0$ in \re{H}. This is because the Baryon charge
density $\varrho_{top}$, is a topological charge density and hence
essentially a total divergence. Thus if one sought conditional solutions
restricted to {\it unit} Baryon number by adding the density
$\xi\varrho_{top}$ to the static
Hamiltonian, with $\xi$ the Lagrange multiplier, then the Euler-Lagrange
equations would remain unchanged, justifying the integration of these
subject to \re{asymf} without it being necessary to have $\la\neq 0$ in
\re{H}. The $\la=0$ case was the problem analysed in \cite{nl,jmp}, and
the solutions constructed there must be considered to be {\it conditional}
solutions. Here we have studied the case woth $\la\neq0$, and found that
switching $\la$ on results in considerable qualitative effects on the
(gauged) Skyrme soliton.

This is not a surprising result, but nor is it predictable. In the present
case $\la\neq 0$ with asymptotics \re{asyma}, it was possible to construct
solutions numerically with both asymptotic values $a(\infty)=\pm 1$. By
contrast in the $\la=0$ case with asymptotics \re{asyma0}, it turned out
\cite{nl,jmp} that out of the three possible asymptotic values \re{asyma0}
only for the two, $a(\infty)=0$ and $1$, could solutions be constructed
numerically. Moreover, the respective energy profiles of these solutions
are quite different as seen from Figs. 1 and 2.

Concerning the phyical relevance of these solitons, first we note thay
they describe the low energy properties of the Nucleons~\cite{ANW}, and
then note that the $SO(3)$ gauging of this system is a perfectly natural
step in the direction of the studying the effects of the
$SO(3)\times SO(2)$ Standard Model on the Nucleons.

\bigskip

\bigskip

\newpage
\begin{small}

\end{small}
\medskip
\medskip

\centerline{Figure Captions}
\begin{itemize}

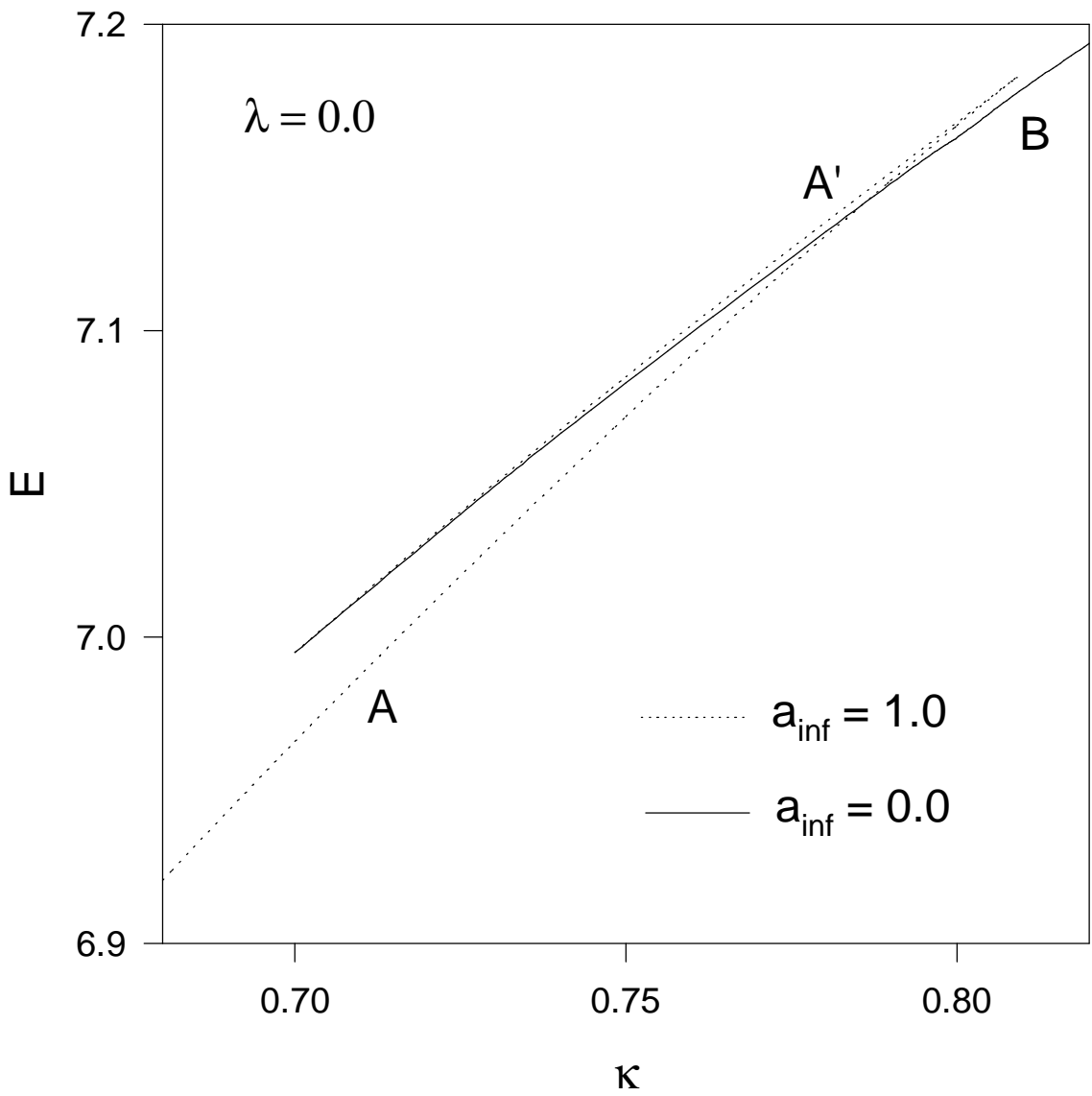
\item [Figure 1]
Energy versus $\ka$ for $a(\infty)=1$ solutions (branches $A$, $A'$)
and for $a(\infty)=0$ solutions (branch $B$), for
(for $\la = 0.0$).

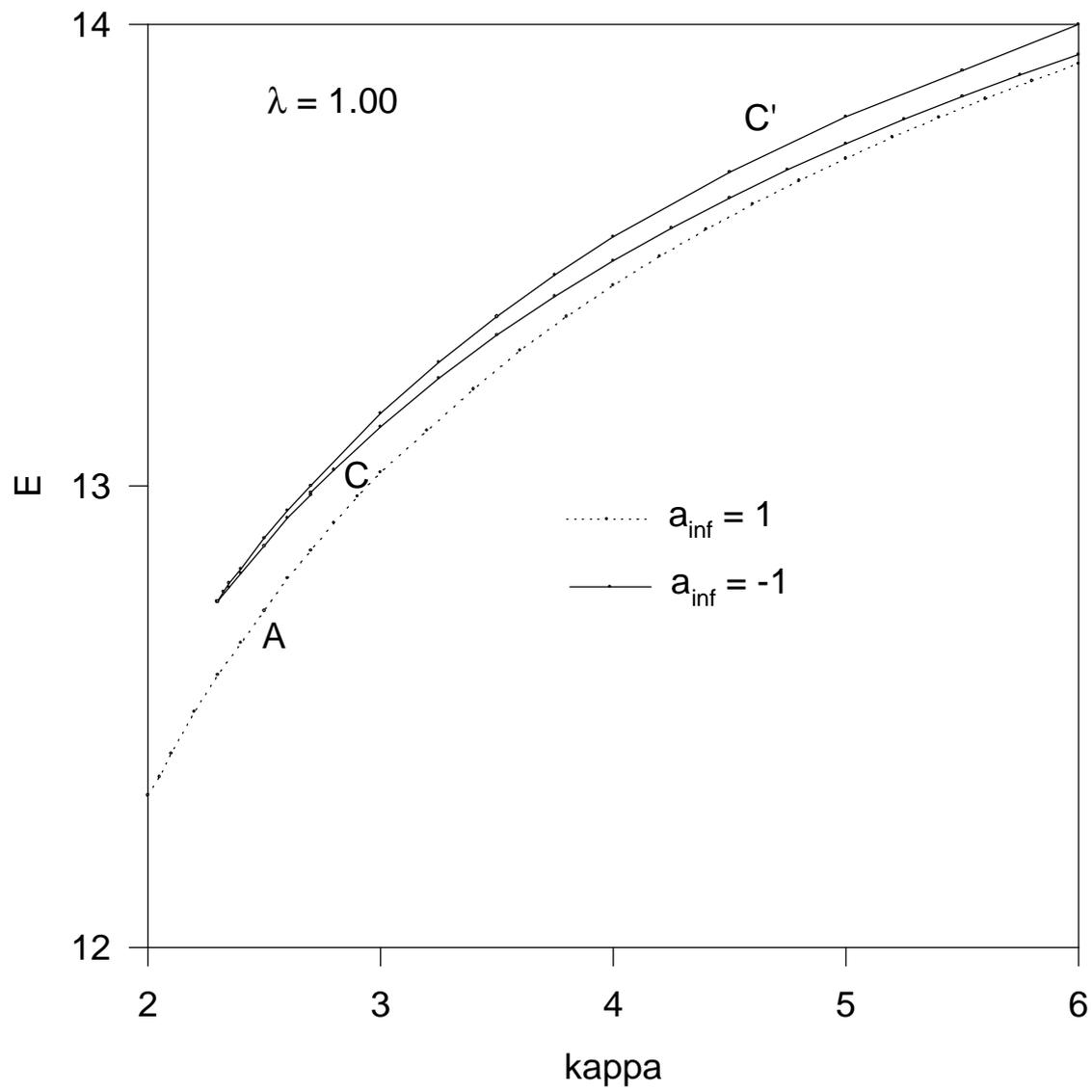
\item [Figure 2]
Energy versus $\ka$ for the $a(\infty)=1$ solution (branch $A$)
and for the $a(\infty) = -1$ solutions (branches $C$,$C'$)
(for  $\la = 1.0$).

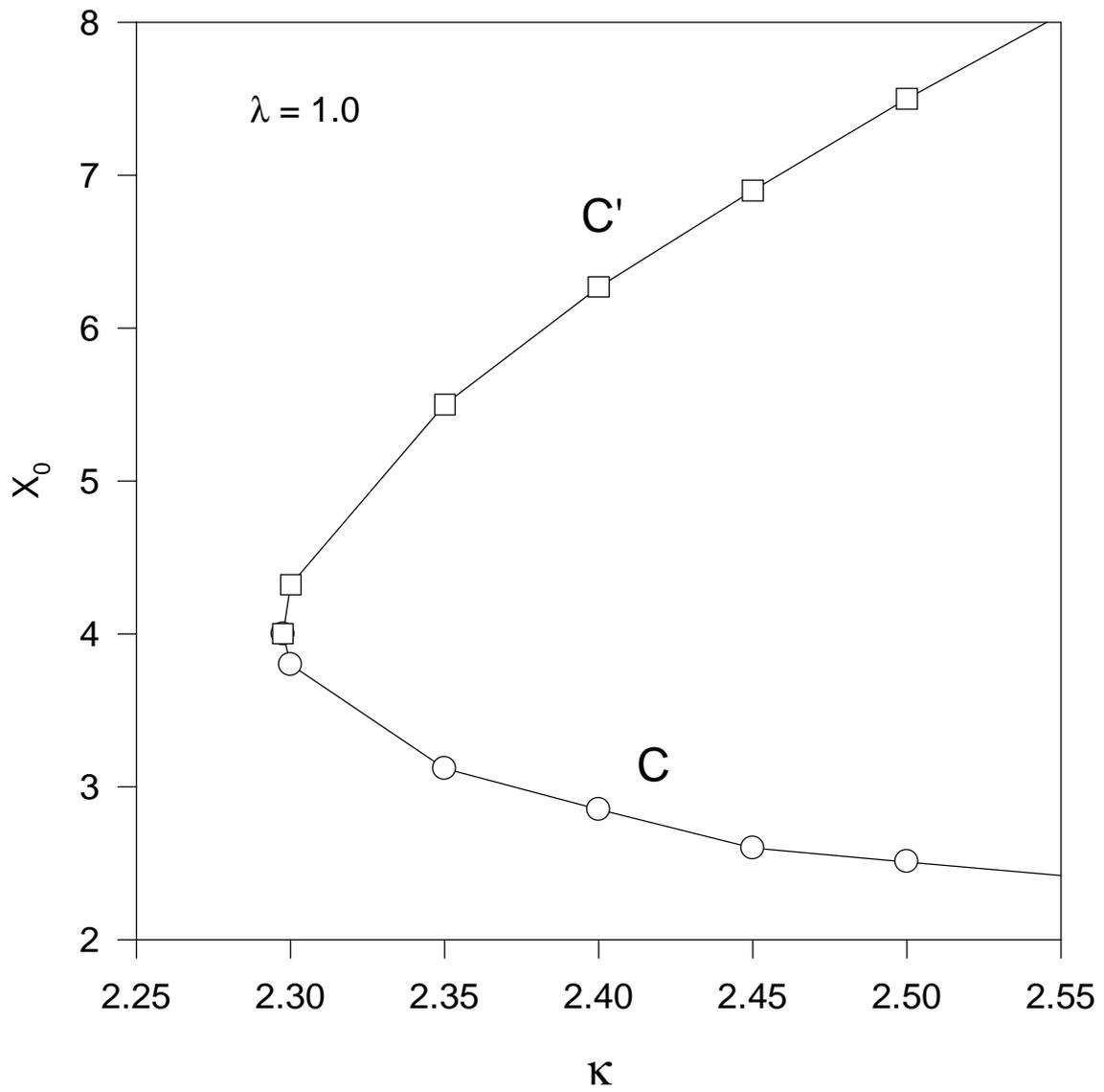
\item [Figure 3]
The evolution of the position of the node of the function $a(x)$
as a function of $\kappa$ for the branches $C$ and $C'$.

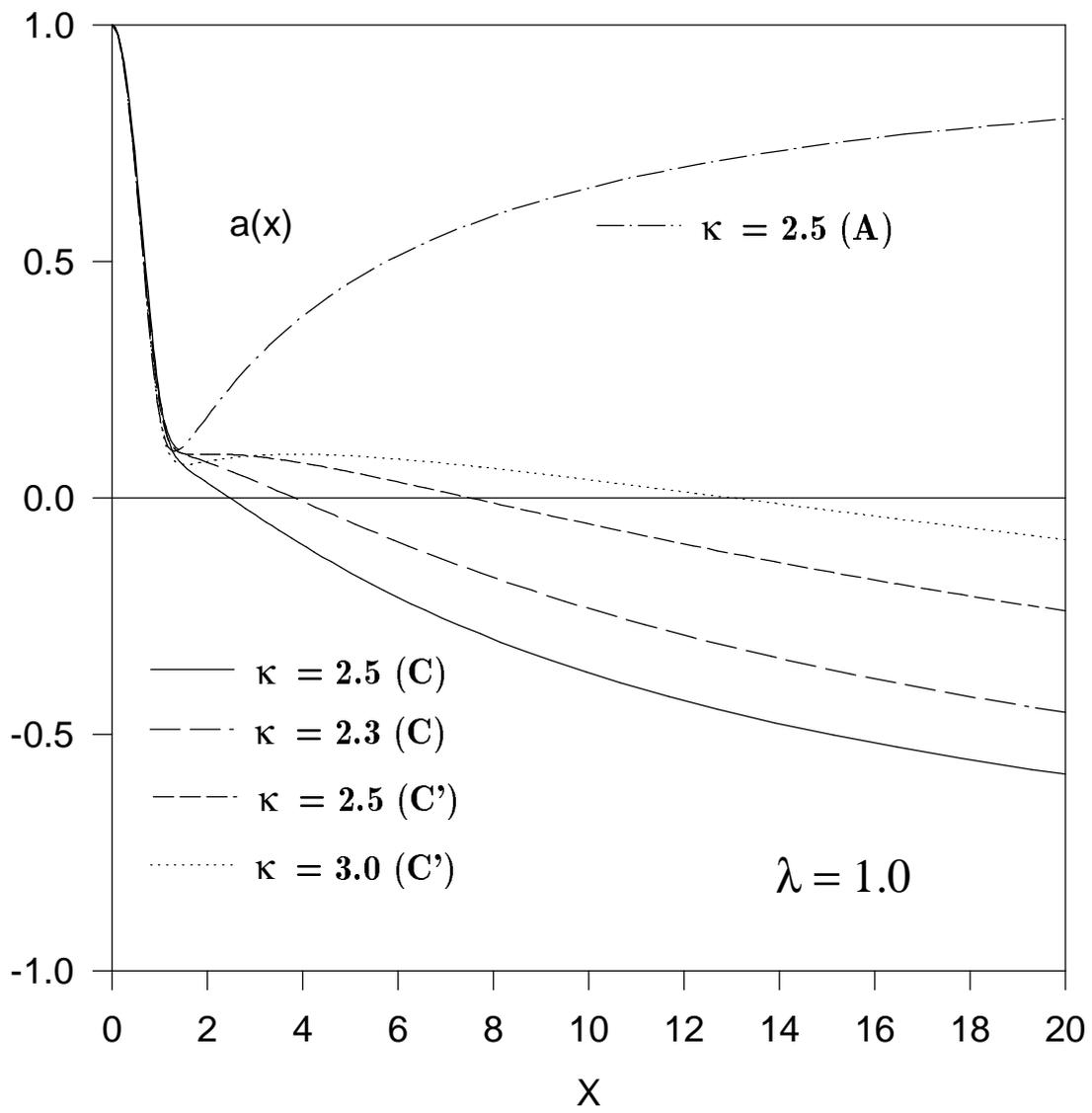
\item [Figure 4]
The profiles of $a(x)$ for the three solutions available for
$\kappa = 2.5$ are presented~: 
$C$ in solid line, $C'$ in short-dashed line, $A$ in dot-dashed line. 
The profiles of $a(x)$ at the critical value ($\kappa \approx 2.3)$
and for $\kappa= 2.5$ on branch $C'$ are supplemented 
respectively on the long-dashed and dotted lines.

\end{itemize}
\end{document}